# Inclusion and Exclusion Criteria in Software Engineering Tertiary Studies: A Systematic Mapping and Emerging Framework

Dolors Costal
Universitat Politècnica de Catalunya
Barcelona, Spain
dolors@essi.upc.edu

Carles Farré
Universitat Politècnica de Catalunya
Barcelona, Spain
farre@essi.upc.edu

Xavier Franch
Universitat Politècnica de Catalunya
Barcelona, Spain
franch@essi.upc.edu

Carme Quer
Universitat Politècnica de Catalunya
Barcelona, Spain
cquer@essi.upc.edu

## ABSTRACT

**Context**: Tertiary studies in software engineering (TS@SE) are widely used to synthesise evidence on a research topic systematically. As part of their protocol, TS@SE define inclusion and exclusion criteria (IC/EC) aimed at selecting those secondary studies (SS) to be included in the analysis. **Aims**: To provide a state of the art on the definition and application of IC/EC in TS@SE, and from the results of this analysis, we outline an emerging framework, TSICEC, to be used by SE researchers. **Method**: To provide the state of the art, we conducted a systematic mapping (SM) combining automatic search and snowballing over the body of SE scientific literature, which led to 50 papers after application of our own IC/EC. The extracted data was synthesised using content analysis. The results were used to define a first version of TSICEC. **Results**: The SM resulted in a coding schema, and a thorough analysis of the selected papers on the basis of this coding. Our TSICEC framework includes guidelines for the definition of IC/EC in TS@SE. **Conclusion**: This paper is a step forward establishing a foundation for researchers in two ways. As authors, understanding the different possibilities to define IC/EC and apply them to select SS. As readers, having an instrument to understand the methodological rigor upon which TS@SE may claim their findings.

## 1 Introduction

A tertiary study is defined as "a systematic review of systematic reviews" [1]. They are widely adopted in software engineering (abbreviated as **TS@SE**), because they offer the possibility to integrate existing knowledge that has been previously synthesised in secondary studies (**SS**) [2]. TS@SE define inclusion and exclusion criteria (**IC/EC**) to select the SS to be included in the review. While their meaning is conceptually simple, a quick look at existing TS@SE reveals disparities at a number of concerns: which conditions appear as IC/EC, how and when they are applied, etc. This paper has two objectives: 1) analyse systematically the state of the art of IC/EC definition and application in TS@SE; 2) based on the results, outline an emerging framework called TSICEC for the definition and application of IC/EC in TS@SE.

## 2 Background

Kitchenham and Charters' guidelines [1] propose three main phases in systematic reviews: *Planning*, *Conducting*, *Reporting*. Inside *Conducting*, the S*election of primary studies* stage is meant to be performed between the *Identification of research* stage and the *Study quality assessment*. Main recommendations for study selection may be summarized as follows:

- Base IC/EC on research questions.
- Pilot IC/EC to ensure that they can be reliably interpreted and they classify studies correctly.
- Apply IC/EC based on practical issues, e.g. [3].
- Maintain a record of those studies that are excluded.
- Measure the level of agreement among the different researchers using Cohen's Kappa statistic [4].
- Discuss and resolve every emerging disagreement.

Fink [3] proposes two consecutive screens to select primary studies: *(i)* "primarily practical", meant to "identify a broad range of potentially useful studies", matching the *Selection* stage in [1]; *(ii)* "for methodological quality", coinciding with the *Quality Assessment* stage in [1]. Besides, Fink [3] provides examples of "practical" IC/EC classified into 13 different *Types*, e.g. *Publication Language* and *Research Design*.

Kuhrmann et al. [5] propose a "practical and pragmatic" guideline for designing systematic studies in SE based on the broad experience of their authors. Regarding the *Selection* stage, the main contribution is twofold.

- A template consisting of a list of "standard" IC/EC that can be conveniently instantiated.
- A comprehensive description of decision guidelines to resolve disagreements and select the papers, with a detailed account of different voting procedures.

## 3 Research Method

Our state of the art on IC/EC in TS@SE takes the form of a systematic mapping (**SM**), following the guidelines proposed by Kitchenham and Charters [1]. To confirm the need for the review, we checked the existence of papers with similar aims. First, we noticed that few contributions, mainly the ones cited in the previous section, address how to properly define IC/EC although, in our opinion, not in a systematic way. Second, no paper is aimed at tertiary studies, i.e., on selecting not primary studies but SS. This significantly differs from our focus on SS because primary studies and SS are fundamentally different: while primary studies are really diverse, SS shall fulfill some methodological criteria well-established in the SE community.

### 3.1 Research Questions

Table 1 defines the research questions (**RQ**) of our SM.

**Table 1. Research questions (RQ) of our study**

| RQ1 | What type of conditions are defined as IC/EC in TS@SE? |
|---|---|
| RQ2 | What smells are observed in the definition of IC/EC in TS@SE? |
| RQ3 | How are IC/EC applied in TS@SE? |
| RQ4 | How are authors involved in the application of IC/EC in TS@SE? |

### 3.2 Search Protocol

In this paper, we combined automatic search in Scopus and backward snowballing as follows:

1. We defined the IC/EC enumerated in Table 2. We fixed 2004 as the starting date because it was the year of publication of the seminal paper on Evidence-Based Software Engineering [6]. The papers to be selected had to fulfill all the IC and none of the EC.
2. IC1 guided the search string definition which after some piloting, ended up as "tertiary study". During the process, we fine-tuned the Scopus search parameters too. IC2 and IC3 were implemented through these parameters, using *Subject Area* and *Publication Year*, respectively. We defined title, keywords and abstract as scope of search.
3. We executed the search string over Scopus with date 19-January-2021, resulting in 103 candidate papers.
4. We applied IC1, IC2 and the EC over title, abstract and keywords of these 103 papers, excluding 46 of them. IC/EC were applied by two team members to every paper, with the agreement not to exclude papers in case of doubt. Only 7 out of the 103 papers were conflictive, but plenary meetings led the team at full consensus.
5. Later on, during data extraction, we excluded 10 works whose full text showed that they did not fulfil some IC or fulfilled some EC. Exclusion was proposed by the team member in charge of extracting the data, and discussed and agreed upon in a plenary meeting.
6. Finally, we performed backward snowballing using as seed the 47 selected papers and we identified 3 additional papers once we applied IC/EC over them.

**Table 2. Inclusion/Exclusion criteria (IC/EC) of our study**

| IC1 | The paper is a tertiary study |
|---|---|
| IC2 | The scope of the paper is software engineering |
| IC3 | The paper is published from 2004 onwards |
| IC4 | The paper is written in English |
| EC1 | The paper is superseded by a later version from the same authors |
| EC2 | The paper describes the TS with very little detail |
| EC3 | The paper is not available (even after contacting authors) |

The 50 selected TS@SE were diverse in different respects, remarkably in their variability in terms of type of SS included. It is worth mentioning that 16 TS@SE did not require SS to be systematic, while 23 required the SS to be either a SLR or a SM. The average number of SS per TS@SE is 75.

### 3.3 Data Extraction, Analysis and Reporting

We stored the search result in a GDrive spreadsheet, which was used in the rest of the study. Selected TS@SE are numbered [S01]-[S50]. We kept track of excluded papers. We added as many columns as needed to extract the data required to answer RQ1-RQ4. One team member was in charge of extracting the data for every paper; the set of papers were split into the team members at equal share. We held weekly plenary meetings to analyse progress and discuss issues as they emerged.

We used content analysis to inductively synthesise codes from the extracted data, grouping them into categories. In general, we paid attention to avoid researcher bias through several actions: working in pairs, supervision (i.e., a researcher validating results from another), weekly plenary meetings and explicit check of inter-rater agreement when necessary.

The complete reporting is available in a replication package [7].

## 4 State of the Art in TS@SE

In this section, we respond to the four RQs. Along the section, we write code values in italics.

## 4.1 RQ1: Conditions defined as IC/EC

We identified 15 IC/EC categories (see Table 3) with a total of 94 codes, being *Contents* the category with more codes (25) and *Peer-reviewed* the one with less (only one). The complete coding schema is detailed in our replication package [7].

**Table 3. IC/EC found in TS@SE**

| Category | Examples (codes in italics) | #IC | #EC |
|---|---|---|---|
| SS_Type | *Generic*, *SLR*, *SM*, *Survey*, ... | 37 | 13 |
| Methodology | E.g., a *Reference* / *Approach* is *followed* | 12 | 12 |
| Quality assessm. | E.g., minimum score applying DARE | 1 | 0 |
| Scope | Typically, *Title+Abstract* or *Full-text* | 4 | 1 |
| Contents | E.g., the SS includes *Lessons-Learned* | 23 | 18 |
| Domain | All *SE*, a *SE-area*, a *Transversal* topic, ... | 35 | 23 |
| Language | Typically, requested to be *English* | 11 | 17 |
| Dates | Publication *After* / *Before* some date | 8 | 2 |
| Authorship | E.g., authors' *Nationality* | 1 | 0 |
| Venue | *Journal*, *Conference*, *Grey-literature*, ... | 17 | 10 |
| Document-type | *Full-paper*, *Short-paper*, *Abstract*, ... | 7 | 22 |
| Availability | Whether the SS found was *Available* | 6 | 9 |
| Peer-reviewed | Whether the SS was *Peer-reviewed* | 20 | 4 |
| Duplicates | A SS is reported in different documents | N/A | 19 |
| Not-ICs | An EC stating that not all ICs are met | 0 | 1 |

Table 3 shows a total of 182 category occurrences in IC and 252 in EC. These numbers arose once we applied our coding. If we sum the actual number of IC/EC declared by authors in their papers, magnitudes decrease: 152 IC and 178 EC. Section 4.2 will help to understand this discrepancy.

**Dominant categories**. These are *Domain* and *SS-type*. For *Domain*, a majority of TS@SE focus on one particular *SE-area* (product lines, requirement patterns, ...) but others are more general and target *SE*, *CS*, ... We find TS@SE which target a topic *Transversal* to SE (typically, teaching). For *SS-type*, sometimes it is completed by a concrete *Methodology* that provides additional details (e.g., "The paper is an SLR, written following the guidelines given in [22]").

**Marginal categories**. We only found one TS@SE defining *Quality-assessment* (**QA**) as IC/EC. We checked whether this was a consequence of having a low number of TS@SE performing QA and this was not the case: up to 35 TS@SE performed QA. *Authorship* was also marginal, even if it is one of the types mentioned by Fink [3] (in the context of the medical domain). Three other categories such as *Dates*, *Availability* and *Scope* are justified because they are often implicit IC/EC.

## 4.2 RQ2: Smells in IC/EC definition

During RQ1's coding process, we systematically collected and then coded those smells that can be problematic (e.g., in terms of understandability or traceability) in IC/EC definition.

**Ill-reporting**. Less than half of the papers (22) report neatly the IC/EC in a *Table* or at least as clearly *Separated-text* with associated *Identifiers*. Ill-reporting makes IC/EC localization and understanding challenging, and sometimes it is not clear e.g. when a condition is being interpreted as IC or as EC.

**Implicit IC/EC**. Even in the cases with optimal reporting, a majority of TS@SE (up to 34) applied additional conditions to filter SS out. We found as main reasons: *(i)* conditions about *Dates*, *Scope* and *Language* are directly implemented in the search stage before defining the IC/EC; *(ii)* conditions concerning *Quality-assessment* and *Duplicates* are considered after selection; *(iii)* the explanation of IC/EC is not clearly delimited in the paper structure (see *Ill-reporting* smell above).

**Undefinition as IC/EC**. Although Table 3 shows that some categories are more prone to appear as IC (e.g., *SS-Type*, *Dates*) and others as EC (e.g., *Duplication*, *Document-Type*), most of them appear as IC or EC indistinctly (e.g., *Availability*, *Language*). Appearing one way or another does not affect the selection outcome, but having a convention to be followed would simplify matters both to TS@SE authors and readers.

**Overloading.** Half of the TS@SE (25) include some IC/EC stating more than one condition. E.g., "Systematic reviews or mapping of the literature in the TD area", mixing *SS-type* and *Domain*. Overloading impacts on *(i)* comprehensibility of individual IC/EC and *(ii)* traceability to keep track of the real cause of exclusion of a SS.

**Spreading**. Conversely, one third of TS@SE (17) stated conditions spread over several IC/EC. E.g., "[one EC] it is a poster, short paper, doctoral symposium paper, theses or dissertation, working papers; [another EC] it is a summary of conferences, editorial, or workshops".

**Verbosity**. A few TS@SE (6) mixed the IC/EC definition with text that justifies or comments the stated condition. E.g., "The paper has been published between 2007, when Kitchenham and Charters [21] proposed their guidelines, and 2016".

**Redundancy**. A low number of TS@SE (6) included conditions that are unnecessary, or stated in different words in the same or different IC/EC. E.g., "The systematic mapping study included a systematic review process" (but all SM do).

**Cross-redundancy**. A dominant case of redundancy (24 TS@SE suffered it) in which some conditions are stated both as IC and EC (in its negative form). An extreme case includes six conditions expressed both as IC and EC (e.g., "[*IC*] Only studies in English; [*EC*] Other languages").

**Ambiguity**. Some terms may have different meanings in different (or even the same) TS@SE. An example is the concept of "duplicate", which sometimes has the meaning of the same document (i.e., same DOI) retrieved more than once, and sometimes has the meaning of the same SS reported in more than one document (i.e., different DOI). The term "grey literature" is also interpreted in different ways in diverse TS@SE.

**Inaccuracy**. A low number of TS@SE (5) state conditions that are not accurate. E.g., "The paper is peer-reviewed (journal article, conference paper)" (workshop papers and book chapters can be peer-reviewed too).

## 4.3 RQ3: Application of IC/EC

The analysis of the papers reporting TS@SE reveals several facets of interest with respect to IC/EC application:

**Phases**. Twenty-one papers report more than one phase when applying IC/EC, with two phases in most of the cases except for two papers that report three phases. It is worth remarking that the frontier among search and the IC/EC-based selection initial phase is sometimes blurry, especially considering IC/EC that can be applied directly in the search, as *Date* or *Source*.

**Scope**. In the 30 papers that report the scope in which the IC/EC were evaluated, *Title+Abstract* (sometimes including *Keywords* too) is the most dominant approach, especially in the first phase of multi-phase TS@SE (17 out of 21). *Full paper* is the second dominant code, used in the last phase in 19 out of the 21 multi-phase TS@SE. In two papers, instead of the full paper, *Introduction+Conclusions* is defined as scope. A pair of papers mention *Metadata* as scope, for those IC/EC that do not require looking at the paper's content (e.g., *Date* or *Language*).

**Application**. Most papers don't report any particular order in applying IC/EC, and we interpret that they apply all of them at once for every candidate SS. But a few TS@SE apply them *Selectively*, i.e. in a given order, being the usual reason that some IC/EC can be assessed by just examining *Title+Abstract* or *Metadata*, while others require *Full paper*. A couple of them apply first all IC and then all EC, or the other way around; for instance, [S26] performed a mix of TS@SE and SS and this is why "In our protocol, we decided to apply the exclusion criteria before inclusion criteria. The main reason for that is because primary and secondary studies shared the same exclusion criteria, but they had different inclusion criteria".

**Interpretation**. Almost all the papers interpret IC/EC as: "To be included, a systematic review needed to meet all of the inclusion criteria, while it could be excluded if it met any of the exclusion criteria" [S23]. Still, we found a couple of cases in which a SS was included if it fulfilled any IC, e.g. "IC1: The study features a systematic literature review; IC2: The study features a systematic literature mapping; IC3: The study features a literature survey" [S10].

**Snowballing**. Twenty TS@SE applied snowballing. We analysed how the IC/EC were applied over the papers found in snowballing and we identified four approaches:

- *Search-IC/EC-snowballing-IC/EC* (2 papers). The selection process applies IC/EC first to the search result, then to snowballing over the result of this first filter.
- *Search-snowballing-IC/EC* (5 papers). Snowballing follows automatic and/or manual search. Then, IC/EC are applied in order to make the final selection. Often, it is not clear how candidate SS were identified during snowballing ("All studies found in the additional venues that were not yet in the pool of selected studies but seemed to be candidates for inclusion were added to the initial pool" [S07]).
- *Snowballing-IC/EC* (4 papers). Snowballing is directly applied to a set of seed papers and the result is filtered using IC/EC, possibly after merging with the result of an automatic and/or manual search done independently.
- *Snowballing-without-IC/EC* (8 papers). The paper does not state explicitly whether IC/EC were applied to papers found in snowballing or not. For instance, in Fig. 3 in [S04] ("Selection process") the "Apply IC/EC" step is followed by another one labelled "Snowballing" only, without further reference in the text about how papers were selected during snowballing.
- *Ad-hoc* (1 papers). [S14], applies an ad-hoc approach tied to the particular type of study.

**Miscellaneous**. There are other concepts appearing in the TS@SE that we do not detail here for lack of space, e.g. validation with gold standard, piloting, and consultation of domain experts in the IC/EC application.

## 4.4 RQ4: Involvement of Authors in IC/EC

Up to 13 TS@SE do not provide any detail about authors' involvement in the application of IC/EC over the candidate SS. For the remaining 37 papers, we checked four facets that we describe below. It is worth noting that only 7 papers reported these four facets altogether [S01, S03, S15, S23, S32, S43, S48].

**IC/EC evaluation** (reported by 33 papers). Details on how researchers evaluate IC/EC over candidate TS@SE:

- *Independent* (19 papers). Several researchers evaluate IC/EC independently. Usually, 2 researchers (which could eventually work in different pairs) were in charge of evaluation, although [S11] and [S37] involved 3 and [S31] 5 researchers. One paper with a multi-phase process (see RQ3), [S46], involved 3 researchers in the first phase and 2 researchers in the second phase.
- *Individual* (2 papers). Only one author evaluates the IC/EC.
- *Consecutive* (3 papers). Three multi-phase TS@SE [S08, S13, S42] report that two researchers worked one after the other, with the second one evaluating only those papers not rejected by the first one.
- *Supervised* (6 papers). In this approach, generally one author evaluates the IC/EC, and another author checks the result. The number of researchers may slightly vary ([S28] does not state the number of authors evaluating the IC/EC, while in [S15], two authors supervised the results). One paper, [S15], reports that only two thirds of the evaluations were supervised.
- *Complex*. Last, a six-author paper [S44] reports the application of a complex workflow to evaluate IC/EC.

**Researchers identity**. Up to 21 papers report this facet, and 13 out of them identify authors *Unequivocally*, either by name or otherwise ("the first author...", "the PhD student...", ...). Other two papers [S04, S10] identify researchers *By-ID* and the correspondence between the researchers' identity and the ID cannot be established. Last, six TS@SE assigned researchers *Random*ly; some of them made different pairings over the dataset to reduce researcher bias.

**Conflict resolution** (reported by 31 papers). We remark: *(i)* four papers reporting IC/EC evaluation strategy do not report conflict resolution; *(ii)* conversely, three papers reporting conflict resolution do not report IC/EC evaluation strategy.

- *Consensus* (16 papers). This strategy seeks solving conflicts by discussion by the team of researchers. One paper [S11] states that consensus discussion was supervised by the senior authors.
- *Rule-based* (6 papers). The authors apply some predefined conflict resolution rules. Two particular cases are: *(i)* conservative, in which a SS is filtered out only if all authors agree on this; *(ii)* restrictive, in which a SS is filtered out if at least one author thinks so.
- *Arbitration* (3 papers). One author hears the arguments of disagreement and then makes the decision.
- *Compound* (3 papers). Consensus is used jointly with rule application [S23, S44] and arbitration [S30].
- *Voting* (1 paper). Finally, only one TS@SE [S07] used voting to resolve conflicts.
- Unclear, inconsistent (2 papers).

**Conflict reporting** (reported by 9 papers). Only 6 papers report the number of conflicts that required resolution, and 3 other papers report the use of Kappa inter-rater agreement.

## 5 Discussion

We design our emerging framework in Section 5.1, and then we identify threats to validity and provide a research agenda.

### 5.1 TSICEC: An Emerging IC/EC Framework

From the results of Section 4, we outline the main points for articulating a framework for the definition and application of IC/EC in TS@SE, which we call TSICEC (Tertiary Studies' IC/EC). TSICEC is still emerging and we aim at consolidating it with respect to a number of factors that may influence, e.g. the number of authors, the target venue and the type of document.

**Catalogue of IC/EC**. We propose to keep all the conditions identified in the answer to RQ1. Even if some have not been much used, they respond to the needs of a particular study. Consolidation and evaluation of TSICEC should confirm at the end if all these conditions are to be kept in the catalogue.

**Definition of IC/EC**. We propose to avoid as much as possible the smells that we have compiled in the answer to RQ2: *(i)* completely avoid implicit IC/EC; *(ii)* minimize overloading and spreading as much as possible (if not completely); *(iii)* justifications for IC/EC are welcome but should be clearly separated from the IC/EC definition; *(iv)* (cross-) redundancy, ambiguities and inaccuracies should be consciously avoided.

**Semantics of IC/EC**. In order to have a clear-cut criterion, we propose to define as IC those conditions that are implemented in the search (through search string and/or digital library configuration), and to define as EC the rest. This implies the strong methodological recommendation of defining IC/EC before the search is effectively conducted. Concerning their application (see RQ3), a SS is to be selected in a TS@SE if and only if: *(i)* it fulfils all the IC; *(ii)* it does not fulfil any EC.

**Application of IC/EC**. Application is driven by the following rules: *(i)* apply IC at search time, using digital libraries engines as much as possible, working over metadata, title, abstract and keywords as required by every type; *(ii)* make a first screening by applying not-metadata-related IC, and all EC, over title, abstract and keywords, over the SS resulting from the search; *(iii)* in case of snowballing, apply both IC and EC (since no search is needed but still some ICs need to be checked, e.g. in reference to dates); *(iv)* during data extraction and even synthesis, allow for late IC/EC exclusions over full text. These rules can be combined in different ways in a particular IC/EC depending e.g. on type of search and number of SS found. Other aspects such as piloting or use of quasi-gold standards should be investigated on a case-by-case basis.

**Involvement of researchers**. We recommend at least two researchers applying IC/EC independently: *(i)* working in different pairs as much as the number of authors allow; *(ii)* piloting the application through analysis of Kappa's inter-rater agreement; *(iii)* rate papers as "in", "out" and "undecided"; *(iv)* hold periodical meetings among the pairs of researchers to detect any conflicts or solve undecided papers; *(v)* hold regular meetings among the full team to share and solve together any remaining conflict, creating a shared understanding.

**Reporting of IC/EC**. In four different parts of the paper and depending on the available space given the document type:

- In the *Research Method* section, include: *(i)* a table with the IC/EC in short form; *(ii)* justifications of any unusual IC/EC, or values of some IC/EC (e.g., justification of starting date), but clearly separated from the condition; *(iii)* a short description of the application protocol.
- In a separated (sub)section, briefly summarize, and discuss any relevant outcome, after applying IC/EC (e.g., high number of papers excluded by some particular IC/EC). This (sub)section should be integrated with the results and analysis of related activities: search, QA, ...
- In the *Threats to Validity* section, briefly summarize the most relevant IC/EC-related threats and describe how the protocol described in the *Research Method* mitigates them
- In the *Replication Package*: *(i)* keep all the papers excluded during selection; *(ii)* for each excluded paper, record the IC/EC that motivated the exclusion; *(iii)* include an ***IC/EC digital signature*** of the TS@SE, using the codes and values proposed in RQ1-RQ4's answers. See our replication package example [7] for this very paper.

## 5.2 Threats to Validity

Our study faces several threats to validity. For the sake of space, we present below the most relevant ones only.

**Internal validity**. As any other literature study, we can miss some TS@SE not found in our search. The fact that we conducted snowballing with Google Scholar mitigates this threat with respect to automatic search only. A second relevant threat is handling incomplete, unstructured or even contradictory information found in the TS@SE papers. To mitigate this threat, we also decided not to conduct some analysis that we had planned in order to avoid arriving at dubious results.

**Construct validity**. Since the casuistic we found in the TS@SE was very diverse, we made some decisions with the aim of simplifying the coding when faced with particular situations. E.g., we found some conditional IC/EC (e.g., "SLRs related to Software Cost/Effort estimation" and "SMs related to Software Cost/Effort estimation and datasets, productivity, sizing techniques") but we decided not to code conditions.

**Conclusion validity**. The emerging framework that we are proposing as conclusion of this study is purely theoretical at this point, based on a subset of the existing TS@SE, and still pending consolidation. Therefore, it needs to be considered as a preliminary, but still (we hope) significant, step towards the final goal of having a shared body of knowledge for IC/EC definition and application in TS@SE.

**External validity**. Even if most of our findings look not particular to TS@SE, we do not claim them to be applicable in any other type of systematic review or research area. This would require additional investigation and in particular, to repeat the study over a different dataset of studies.

## 5.3 Research Agenda

We outline here the future steps of this emerging research:

- Include those aspects not reported in this paper for lack of space (we have mentioned some along the paper).
- Complete the SS dataset by: *(i)* repeating the automated search in a more recent date, including other typical digital libraries; *(ii)* updating backward snowballing and conducting forward snowballing.
- Update quantitative results and check validity of observations with the extended dataset.
- Refine the TSICEC framework as follow-up of the new observations and consolidation of existing knowledge.
- Connect this work with other aspects relevant to TS@SE with the ultimate goal of having a complete state of the art and framework for conducting TS@SE. One of these such relevant aspects is the Quality Assessment of SS in TS@SE, for which we have also formulated an emerging framework, QUASY [8].

## 6 Conclusions

In this paper, we have first conducted a systematic mapping to uncover the state of the art in the definition and application of IC/EC in TS@SE with the following main contributions:

- Categorization of the IC/EC defined in TS@SE.
- Identification of the main types of smells found in the literature when defining IC/EC in TS@SE.
- Identification and categorization of the main facets related to the process of application of IC/EC in TS@SE.
- Identification and categorization of the different situations found in authors' involvement.

Building upon these results, we have outlined the TSICEC framework providing practical guidance for authors in different respects. We highlight the need of considering contextual factors that may influence some aspects of the conduction, and the proposal of the concept of IC/EC digital signature to report in a systematic manner the IC/EC-based selection process. Evaluation of this framework (e.g., interviews with authors of the selected papers or focus groups with some of them) is also part of future work.

We have tried to apply the framework to our own paper as much as possible. However, some recommendations arose too late in the study (e.g., the classification done for our IC/EC is not compliant with our recommendation of defining search parameters as IC). In relation to reporting, given the short length of this document, we had also to sacrifice some of the recommendations; however, the paper's IC/EC digital signature included in [7] serves as a complement for this lack of space (which is one of the main purposes of this instrument).

## Acknowledgments

This work has been partially supported by the DOGO4ML Spanish research project (ref. PID2020-117191RB-I00).